\documentclass[12pt,preprint]{aastex}

\shorttitle{Disk Angular Velocity Modified by MRI}
\shortauthors{Kato et al.}

\begin{document}

\title{Modification of Angular Velocity by Inhomogeneous MRI Growth
in Protoplanetary Disks}

\author{M. T. Kato, K. Nakamura, R. Tandokoro}
\affil{Department of Earth and Planetary Science, Tokyo Institute of Technology, Ookayama 2-1-12-I2-10, Meguro-ku, Tokyo}
\email{marikok@geo.titech.ac.jp}

\author{M. Fujimoto}
\affil{Institute of Space and Astronomical Science, Japan Aerospace Exploration Agency, Yoshinodai 3-1-1, Sagamihara, Kanagawa}

\and

\author{S. Ida}
\affil{Department of Earth and Planetary Science, Tokyo Institute of Technology, Ookayama 2-1-12-I2-10, Meguro-ku, Tokyo}

\begin{abstract}
We have investigated evolution of magneto-rotational instability (MRI) in protoplanetary disks that have radially non-uniform magnetic field such that stable and unstable regions coexist initially,
and found that a zone in which the disk gas rotates 
with a super-Keplerian velocity emerges as a result of 
the non-uniformly growing MRI turbulence. 
We have carried out two-dimensional resistive MHD simulations 
with a shearing box model. 
We found that if the spatially averaged magnetic Reynolds number, which is determined by widths of the stable and unstable regions in the initial conditions and values of the resistivity, is smaller than unity, the original Keplerian shear flow is transformed to the quasi-steady flow
such that more flattened (rigid-rotation in extreme cases) velocity profile emerges locally and the outer part of the profile tends to be super-Keplerian.
Angular momentum and mass transfer due to temporally generated MRI turbulence 
in the initially unstable region is responsible for the transformation.
In the local super-Keplerian region,
migrations due to aerodynamic gas drag and 
tidal interaction with disk gas are reversed.
The simulation setting corresponds to the regions near 
the outer and inner edges of a global MRI dead zone in a disk. 
Therefore, the outer edge of dead zone, as well as the inner edge,
would be a favorable site to accumulate dust particles to 
form planetesimals and retain planetary embryos against type I migration.
\end{abstract}

\keywords{accretion disks --- instabilities --- 
MHD --- planetary systems: formation --- turbulence}

\section{Introduction}

The ubiquity of extrasolar planets strongly suggests that planet formation is a common process associated with
star formation.
However, two serious barriers are recognized in 
planet formation theory: meter-size and
type I migration barriers.
Terrestrial planets or icy cores for
gas giants that are embedded in protoplanetary
disks tend to lose orbital angular momentum and 
migrate inward through tidal interaction with the disk gas
(``type I migration'').
Linear calculations \citep[e.g.,][]{tan02}
predict that the planets spiral into the host stars
on timescales $\la 10^5$ years for the minimum-mass solar nebula model, 
which is much shorter than observationally inferred disk lifetime. 
This is the type I migration barrier for survival of
planets with an Earth mass or more.

The meter-size barrier is the barrier for formation of
planetesimals.
Since rotation speed of disk gas is slightly slower
than dust grains or planetesimals due to negative
radial pressure gradient due to global structure of the gas disk
and the motions of meter-sized bodies (boulders) 
are marginally coupled 
to gas motion through aerodynamical gas drag, 
the meter-sized boulders suffer
the fastest inward orbital migration 
and the timescale to spiral into the protostar is 
$\sim$ a hundred orbits \citep{adachi76,wei77}.
This timescale would be too short compared to the sticking timescale
for the boulders to form planetesimals of 1--10 km radius or more,
the motions of which are decoupled from the gas motion, since
the meter-sized boulders are expected to 
stick together only poorly \citep{benz00}. 

One way to bypass the meter-size barrier is formation of
clumps through self-gravitational instability of a dust layer,
which can occur on orbital periods \citep{saf69,gold73}.  
However, local Kelvin-Helmholtz instability due to
difference in rotation velocities between the dust-rich layer 
(Keplerian) and
an overlaying dust-poor layer (sub-Keplerian) 
would prevent dust grains
from settling to the midplane to form the thin enough layer for 
the self-gravitational instability
\citep{weiden93}.
Furthermore, if global turbulence exists in the disk,
it also stirs up the dust grains from the midplane.
Commonly observed strong H$\alpha$ emission from young stars 
implies protoplanetary disks are viscously evolving
accretion disks with turbulent viscosity.
One of the most likely sources for the turbulence is   
magnetorotational instability (MRI) in weakly ionized disk gas
\citep{bal91}.

Although turbulence inhibits formation of the thin dust
layer near the midplane, dust grains could be trapped
into turbulent eddies.
It is expected that meter-sized boulders are concentrated
in anticyclonic eddies \citep[e.g.,][]{barge95,chav00,joha04,ina06}.
In particular, \citet{joha07} carried out a simulation of
evolution of self-gravitating boulders in an MRI turbulent disk
and found that 1000km-sized clumps are formed in eddies.
One of the biggest issues in this model is 
lifetime of eddies.
To form clumps, the eddies must persist until boulders
are accumulated into the eddies and produce highly dust-rich regions. 
Since rapid formation of highly dust-rich regions is required,
\citet{joha07} assumed meter-sized boulders that show 
the most favorable behavior in the disks. 

Trapping of dust grains is also possible at a
locally high-pressure disk radius, because a positive radial pressure gradient 
induces super-Keplerian gas flow in which dust grains suffer tail winds,
while a normal negative gradient induces head winds
\citep[e.g.,][]{naka86,kla05}.
\citet{rice04} demonstrated that meter-sized boulders
can be accumulated in high-density spiral arms of
self-gravitating disks.

The pressure maximum also exists near the inner boundary of
MRI "dead" zone.
Innermost disk regions are thermally ionized, so that
MRI is active there.
In outer regions, X rays from host stars and cosmic rays 
penetrate all the way to the midplane of the disk, 
and the ionization degree may be high enough to activate MRI.
On the other hand, in the intermediate regions, only surface layers are
MRI active and the regions near the midplane may be inactive ("dead") 
\citep{gam96,sano00}.
Assuming steady gas accretion through a disk, disk gas surface density
jumps up at the inner boundary of the dead zone
according to the change in effective viscosity due to turbulence,
so pressure maximum emerges there.

The pressure maximum, in other words, the outer boundary of
a local super-Keplerian region
also halts inward type I migration \citep{tan02,mass06}.
If residual and/or secondary-generated dust grains maintain 
the dead zone when accretion of planetary embryos proceeds, 
the inner boundary of the dead zone is a favorable
cite to retain planetary embryos.
However, the inner boundary is generally located
well inside 1AU.
It cannot provide building blocks for terrestrial 
planets in habitable zones and icy planets nor retain the cores 
to form gas giants around solar-type stars, 
although it would play an important role in
architecture of short-period extrasolar planets.

\citet{kret07}, however, pointed out that
a local dead zone can appear near the ice line and
the ice line can be a spatial barrier for dust migration due to gas drag.
Because dust grains are the most efficient argents for charge recombination,
ionization rates and thickness of the active layers 
rapidly decreases across the ice line due to condensation of
icy dust grains.
Correspondingly, in a range of disk accretion rates,
a local dead zone appears near the ice line and
at the outer boundary of the dead zone large amount of icy grains 
can be accumulated and cores are retained to form gas giants.

\citet{brau08} performed simulations of dust settling and
coagulation with radial migration due to gas drag
and the ice line effect.
They found that the dust to gas ratio increases and formation of
planetesimals may be efficient near the ice line.
\citet{idalin08a} showed that if the ionization rate is
order of magnitude larger than that predicted by
\citet{kret07} due to dust growth [\citet{kret07} assumed
that all the dust grains have $\mu$m sizes],
cores stop migration at the ice line and 
they efficiently form gas giants without significant reduction
of type I migration speed.
Note that in order for the ice line effect to
actually work, disk accretion rate and dust population
must be within some ranges of parameters such that
the thickness of the active layer is comparable to 
that of dead zone near the ice line, although the ranges are not
too restricted.

We here show through MHD simulations
that the pressure maximum associated with
quasi-steady local super-Keplerian rotation may be
created in the MRI marginal regions, such as the
outer boundary of the global dead zone as well as its inner boundary, 
without requirements of persistent turbulent eddies nor 
the ice line effect.
The accumulation of dust grains and retention of planetary
embryos at the outer boundary of the dead zone have
a great importance for formation of terrestrial and jovian planets.
MRI is controlled by the vertical component of magnetic
field ($B_z$) penetrating the disks as well as by ionization degree 
\citep[e.g.,][]{sano00}.
As shown in later sections,
non-uniform temporal MRI turbulence that occurs 
in the marginally stable regions 
transforms disk gas flow into quasi-steady flow that
has local rigid-rotation regions. 
This flow pattern is sustained by non-uniform pressure gradient 
produced by mass transfer associated with the temporal turbulence.
In the outer regions of the local rigid-rotation regions,
gas rotation is super-Keplerian.

Since dead zones may shrink as dust grains grow
and surface areas for charge recombination decrease \citep{sano00},
such marginal state sweeps from outer disk regions to inner regions.
Furthermore, the density fluctuations due to MRI turbulence
induced by the enhanced ionization degree lead to disruptive
collisions of small planetesimals and the collisions
reproduce dust grains \citep{ida08}.
The grains lower the ionization degree so that
disk gas becomes marginally stable against MRI.
Then, grain growth starts again.
Thus, marginally MRI stable state could be maintained
in significant regions at $\la 10$AU until disk gas
is depleted to some degree \citep{ida08}.
\citet{oishi07} showed the random torques due to
the density fluctuations in the surface active layers
affect planetesimals in the dead zone near the disk midplane. 
This effect may result in more continuous dust production
and help maintenance of the marginally stable state.
Thus, it is expected that broad regions at $\la 10$AU
may once experience such local super-Keplerian motions 
and accumulate dust grains to form planetesimals
during disk evolution.

In section \ref {numerical}, we describe the numerical method 
and the initial conditions.  We use a local shearing box with 
non-uniform $B_z$.
In section 3.1, the results with a fiducial model are shown.
In section 3.2, the dependence of the results on initial settings
is presented.
Summarizing the numerical results, we show in section 3.3 
that creation of the rigid rotation and super-Keplerian regions 
is regulated by spatially averaged magnetic Reynolds number.
Section 4 is conclusion and discussion.

\section{Numerical Model}
\label{numerical}

\subsection{Shearing box model}

We carry out local two-dimensional MHD simulations 
of protoplanetary disks in the shearing box model
\citep{wisdom88,haw95}. 
The coordinates are centered at $r_{0}$ from a host star
and rotating with Keplerian angular velocity at $r_0$ ($\Omega_{0}$). 
The radial, azimuthal and vertical directions are $x$, $y$ and $z$,
respectively. 
Assuming uniformity in the $y$-direction,
we simulate 2D flow in the $x$-$z$ plane.
From the flow in the $x$-$z$ plane, the evolution of $v_y$ 
is calculated.
We will discuss coordinate sizes in the $x$-$z$ directions 
($L_x$ and $L_z$) later.
Periodic boundary conditions are applied for the $x$ and $z$ directions.
In the $x$ direction, Keplerian shear motion in the $y$-direction
is taken into account in the boundary condition.

\subsection{Basic equations}

We adopt compressible resistive magnetohydrodynamic (MHD) equations:
\begin{eqnarray}
\frac{\partial \mathbf{v}}{\partial t}
+\left(\mathbf{v}\cdot\nabla\right)\mathbf{v}&=&
-\frac{1}{\rho}\nabla\left(P+\frac{\mathbf{B}^2}{8\pi}\right)
+\frac{1}{4\pi\rho}\left(\mathbf{B}\cdot\nabla\right)\mathbf{B}
-2\mathbf{\Omega}_{0}\times\mathbf{v}+3\Omega^2_{0}x\hat{\mathbf{x}}, \label{motion}\\
\frac{\partial \rho}{\partial t}
+\nabla\cdot\left(\rho\mathbf{v}\right)&=&0, \label{continuity}\\
\frac{\partial \mathbf{B}}{\partial t} &=&
\nabla\times\left[\left(\mathbf{v}\times\mathbf{B}\right)-
\eta\nabla\times\mathbf{B} \right], 
\label{induction}\\
P&=&c^2_{s}\rho, \label{isothermal}
\end{eqnarray}
where $\mathbf{v}$, $P$, $\rho$ and $c_{s}$ are
velocity in the rotating frame, pressure, density and sound speed 
of the gas, respectively.
The third and forth terms in r.h.s. of equation~(\ref{motion})
are the Coriolis force and tidal force (the difference between
the centrifugal force and gravitational force from the central star),
in which $\hat{\mathbf{x}}$ is a unit vector in the $x$-direction.
$\mathbf{B}$ and $\eta$ are magnetic field and resistivity.
For simplicity, we omit the vertical component of the gravity.
We include ohmic dissipation in equation~(\ref{induction}),
while we neglect ambipolar diffusion.
Its effect is weaker than the ohmic dissipation
in midplane of inner ($\la 100$AU) disk regions \citep{jin96},
although in the surface layer or disk inner edge,
dissipation due to ambipolar diffusion may be important \citep{chi07}.

We scale length and time by disk scale height 
($H=c_{s}/\Omega_{0}$) and $\Omega_{0}^{-1}$, respectively.
Then, the independent parameters in the equations 
are plasma beta ($\beta$)
and magnetic Reynolds number ($R_{\rm m}$)
defined respectively by
   \begin{eqnarray}
   \beta = \frac{2 c_s^2}{v_{\rm A}^2}, \label{eq:plasma_beta}\\
   R_{\rm m} = \frac{v_{{\rm A}z}^2}{\eta \Omega_0} \label{eq:mag_Re},
   \end{eqnarray}
where $v_{\rm A} = B/\sqrt{4 \pi \rho}$ is Alfven velocity
and $v_{{\rm A}z}$ is its $z$-component.
We assume spatially uniform and time-independent $c_s$ and $\eta$.

We solve the resistive MHD equations 
using the CIP-MOCCT method \citep{yab91,sto92} with
grid sizes of $0.01H$ (for dependence of our results on resolution, 
see Appendix 1).
We usually integrate the evolution until $t \simeq 100 \Omega_0^{-1}$.
Although most of numerical studies on nonlinear stages of
MRI have assumed ideal MHD, we consider low ionization state 
with finite resistivity.
Several numerical simulations \citep{fle00, haw96} showed 
that the finite magnetic resistivity reduces growth rates of the MRI.

\subsection{Initial conditions}

We assume that gas density is initially uniform ($\rho_0$),
so pressure is also uniform ($P_0$) due to the assumption of 
isothermal gas (constant $c_s$).
We also assign the initial value of 
plasma beta as $\beta = 400$ uniformly.
According to the large $\beta$, we set initial steady flow
as a uniform Keplerian shear flow, $v_{y}=-(3/2) x \Omega_0$.

The remaining parameter is $R_{\rm m}$.
This value determines the MRI growth rate \citep{jin96, sam99}.
For $R_{\rm m} \la 1$, short wavelength modes are stabilized
by ohmic dissipation and the growth rate declines.
For $\Omega_0 \propto r^{-3/2}$, 
the the most unstable wavelength is \citep{sam99}
   \begin{equation}
   \lambda_{\rm m.u.} \simeq 2 \pi \frac{\eta}{v_{{\rm A}z}}
   \simeq \frac{2\sqrt{2}\pi}{\sqrt{\beta} R_{\rm m}}H 
   \simeq \frac{0.44}{R_{\rm m}}H. 
   \label{eq:lambda.m.u}
   \end {equation}
Since MRI would not occur in the disk if 
$H \la \lambda_{\rm m.u.}$, the stabilization 
condition is $R_{\rm m} \la R_{\rm m,crit} \simeq 0.44$.

We consider marginally stable state for MRI,
in which stable and unstable regions co-exist non-uniformly.
In this study, we assume constant $\eta$, and
the non-uniformity of $R_{\rm m}$ is set by that of
$v_{{\rm A}z}$ (equivalently, $B_z$).
Since non-uniform MRI is an essential point for 
the emergence of the super-Keplerian flow, 
similar results are obtained in the case of non-uniform $\eta$.
We will show the results with non-uniform $\eta$ in a next paper.

\begin{figure}[tbp]
\begin{center}
\includegraphics[width=100mm]{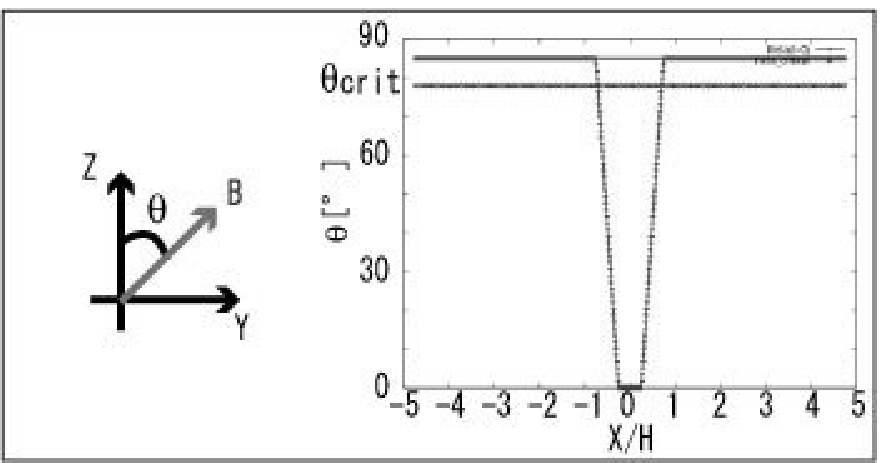}
\end{center}
\caption{The configuration of the initial magnetic field. 
$\theta$ represents the angle from the vertical direction ($z$) 
to the azimuthal one ($y$).
It radially varies as shown in the right panel.
With constant $B_0$, the initial magnetic field 
is given by $\mathbf{B}=(0, B_{0}\sin\theta, B_{0}\cos\theta)$. 
For $\eta = 0.002 H^2 \Omega_0$,
linear calculations suggest that MRI occurs for 
$\theta < \theta_{\rm crit} \simeq 79$ degrees.}
\label{fig:initial}
\end{figure}

In order to represent the non-uniformity of $B_z$ with uniform $\beta$, 
we set initial $\mathbf{B}_0$ such that
   \begin{equation}
   \mathbf{B}_0=(0, B_{0}\sin\theta, B_{0}\cos\theta) 
   \label{eq:initial_B}
   \end{equation}
with uniform $B_{0}$ and non-uniform $\theta$.
As shown in Figure \ref{fig:initial}, we set
$\theta = 0$ ($\cos \theta = 1$) in the middle region (unstable)
with radial size $L_{\rm u}$
and $\theta = 85$ degrees ($\cos \theta \simeq 0.087$) 
in the side regions (stable) with individual radial size $L_{\rm s}$.
The box size in the horizontal direction is given by
$L_x = L_{\rm u} + 2 L_{\rm s}$.
Transition zones between $\theta = 0$ and 
85 degrees are set to avoid numerical instability.
We include non-zero azimuthal magnetic component $B_y$
because the plasma beta and therefore the magnetic pressure are 
set to be spatially uniform to establish the initial equilibration. 
The azimuthal component $B_y$ is calculated even in these two-dimensional 
simulations on the $x$-$z$ plane, but the assumption of axisymmetry excludes 
the occurrence of magnetic dynamo on the $x-y$ plane due to
reconnection of $B_y$ and the results here do not change 
even if $B_y$ is set to be zero.
In our preliminary three-dimensional simulations,
we find that the MRI growth due to reconnection of $B_y$
in the side areas hardly affects the features 
of azimuthally averaged fluid motion.
This is because the essential process 
of the transformation from Keplerian flow to
quasi-steady non-uniform rotation flow 
is temporal MRI growth and the stabilization of MRI 
due to established rigid rotation
but not due to dynamo of magnetic field (see section 3).
The effects of three-dimensional flow 
will be addressed in detail in the next paper.

The above choice indeed situates both stable and unstable regions 
in the simulation box.
Substituting equations~(\ref{eq:plasma_beta}) and (\ref{eq:initial_B})
into equation~(\ref{eq:mag_Re}), we obtain
   \begin{equation}
   R_{\rm m} = \frac{v_{\rm A}^2 \cos^2 \theta}{\eta \Omega_0}
             = \frac{2 \beta^{-1} H^2 \Omega_0 \cos^2 \theta}{\eta}
             = 2.5 \left( \frac{\eta}{0.002 H^2 \Omega_0}\right)^{-1} \cos^2 \theta.
   \label{eq:mag_Re_fudcial}
   \end{equation}
With a fiducial value $\eta=0.002H^{2}\Omega_0$, MRI is
initially activated ($R_{\rm m} > R_{\rm m,crit}$)
in the middle region with
$\theta < \theta_{\rm crit} \simeq 65$ degrees.
The exact dispersion relation
(Appendix 2) show that 
wavelengths of all the growing modes exceed the vertical box size $L_z$
for $\theta < \theta_{\rm crit} \simeq 79$ degrees.
The vertical size $L_z$ is set to be $\simeq 0.5 H$
that is comparable to
the most unstable wavelength $\lambda_{\rm m.u.}$
(equation~[\ref{eq:lambda.m.u}]).

We have carried out 79 runs in total with various
$\eta$, $L_{\rm u}$, and $L_{\rm s}$.
The detailed simulation parameters are given in Table \ref{tab:1}.
In some cases, we adopted larger values of $\eta$,
in which $\lambda_{\rm m.u.}$ is larger
(equation~[\ref{eq:lambda.m.u}]).
Then, we adopt $L_z$ as large as $\lambda_{\rm m.u.}$ with
$R_{\rm m} \sim 1$.
In all runs,
random fluctuations are initially given in the velocity field
with a maximum amplitude of $\left|\delta v_{x}\right|=0.001c_{s}$.

\section{Results}
\subsection{The Fiducial Model}\label{modelA}

We first present the detailed results from a fiducial model with
$\eta=0.002H^{2}\Omega_0$, $L_{\rm u} = 1.43H$, 
$L_{\rm s} = 4.05H$, and $L_z = 0.5H$.
Compared with the other models, this model initially has the 
largest stable regions.
If the whole region has uniform and strong enough
magnetic field, MRI turbulence is self-sustaining.
However, evolution of turbulence is quite different
in the case of non-uniform magnetic field that we consider here.
MRI turbulence appears only tentatively, but velocity field
is transformed from Keplerian shear flow to another quasi-steady flow.  

\begin{figure}[tbp]
\begin{center}
\includegraphics[width=100mm]{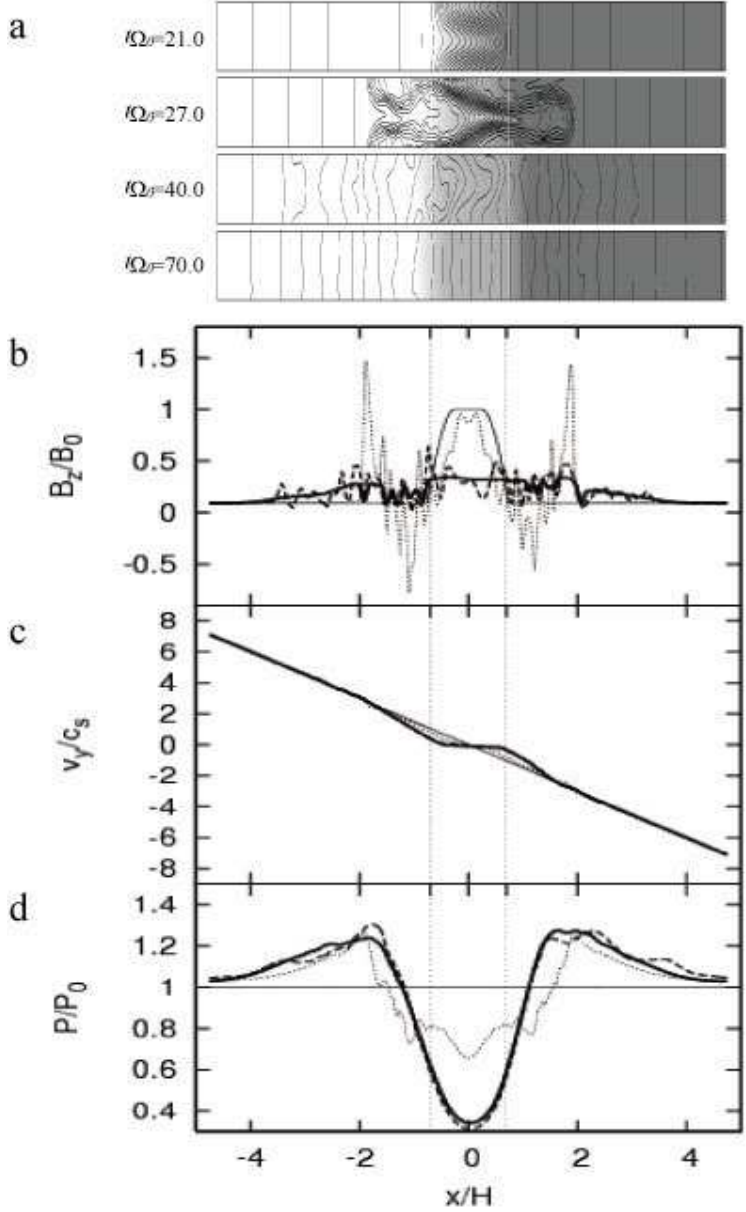}
\end{center}
\caption{Results of the fiducial model. 
(a) Time evolution of the magnetic field (black lines) and 
angular velocity $v_y$ (contours) on the $x$-$z$ plane.
The maximum (brightest) or minimum (darkest) of the tone
on the contours are set at the unstable and stable region boundaries.
Time evolution of vertically averaged values of
(b) the vertical magnetic component $B_z$,
(c) angular velocity $v_y$,
and (d) pressure $P$, as functions of $x$.
In panels (b), (c) and (d), 
thin solid, dotted, dashed and bold lines express
the snapshots at $t\Omega_{0} = 0.0, 27.0, 40.0$ and $70.0$,
respectively.
The unstable region is initially set between the two 
vertical dotted-lines. 
}
\label{fiducial}
\end{figure}

Time evolution of the magnetic field on the $x$-$z$ plane
is shown in Figure~\ref{fiducial}a.
In the panel of $t\Omega_{0}=21$, the field lines 
start to be stretched only in the initially unstable region. 
At $t\Omega_{0}=27$, MRI turbulence develops and
the turbulence is transported into the initially stable regions.
Accordingly, the magnetic perturbations become weaker and 
eventually vanish before they reach the boundary
of the computing box 
(before they fully fill the stable region). 

Figure~\ref{fiducial}b shows evolution of $B_z$ averaged over $z$.
Turbulent diffusion smooths out $B_z$, however,
its level does not go below the critical value for stabilization of MRI 
($\sim 0.2 B_0$) in the initially unstable region 
(the middle region) even at $t\Omega_{0}=70$.
Nevertheless, MRI turbulence ceases after $t\Omega_{0} \ga 40$.
This is because in the middle region,
rigid rotation is realized 
as a result of the angular momentum transfer 
during the MRI turbulence there (Fig.~\ref{fiducial}c) and 
the free-energy source for MRI (differential rotation) is lost.
In the other areas, $B_z$ is smaller than the critical value
and MRI does not grow.
The stabilization due to rigid rotation is discussed 
with linear analysis in Appendix 2.

As shown in Figure~\ref{fiducial}a, the region of the rigid rotation
expands as propagation of turbulence.
This local rigid rotation is self-sustaining, at least,
until the remaining $B_z$ is diffused out by ohmic dissipation
on timescale $\sim L_u^2/\eta \sim 10^3 \Omega_0$ 
(see discussion in section 4). As long as certain level of $B_z$ is maintained, 
deviation from the rigid rotation excites MRI turbulence 
and this transfers angular momentum to bring the velocity profile back to the rigid rotation state.

\begin{figure}[tbp]
\begin{center}
\includegraphics[width=80mm]{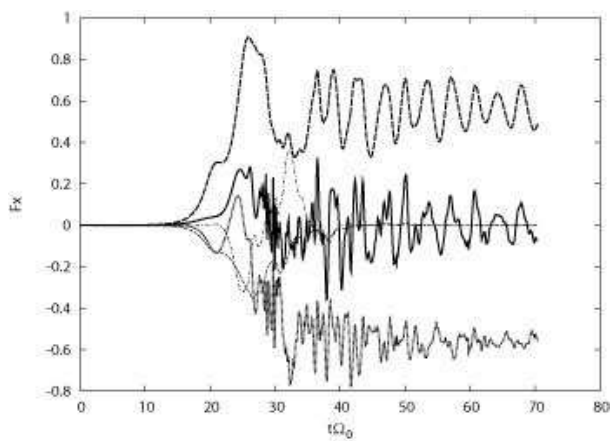}
\end{center}
\caption{Time variation of forces exerted on fluid at $x/H=0.7$ and $z/H=0.25$,
at which super-Keplerian flow is established.
The dotted, dash-dotted, thin dashed and dashed lines express gas radial components of 
pressure gradient, 
magnetic pressure gradient, 
magnetic tension, sum of gravity and Coriolis forces
in the r.h.s. of the equation of motion (equation~[\ref{motion}]), respectively.
The thick solid line is total force.
The unit of the forces is $H \Omega_0^2$.
The pressure gradient and the gravity/Coriolis forces
are dominated and approximately equilibrate with each other. 
}
\label{force}
\end{figure}

Figure \ref{fiducial}d is the vertically averaged pressure distribution. 
The MRI turbulence radially transports mass, associated with
the angular momentum transfer.
Since in the unstable region, mass transfer is efficient
and gas density is proportional to pressure in our isothermal model,
pressure is decreased in the unstable region
while it is increased in the stable regions
adjacent to the unstable region.
The resultant pressure gradients near the boundaries between
the stable and unstable regions equilibrates with 
the Coriolis and tidal forces raised by 
a deviation from Keplerian rotation.
In Figure~\ref{force}, we plot radial components
of individual terms in the r.h.s. of 
the equation of motion (equation~[\ref{motion}])
near the right boundary between the stable and unstable regions 
($x/H=0.7$, $z/H=0.25$). 
At $t \Omega_0 = 20$--40, the amplitude of magnetic pressure 
and tension are increased by the development of MRI turbulence 
and gas pressure gradient is also raised
according to mass transfer associated with the turbulence.
After the MRI turbulence ceases, the pressure gradient 
almost equilibrates with the Coriolis and tidal forces. 
Thus, the rigid rotation with pressure variation caused by the
MRI turbulence is quasi-steady.

As stated in section 1, dust grains can be trapped in
outer edge of local super-Keplerian regions, because
dust grains suffer tail wind and migration outward 
in the super-Keplerian regions (see sections 1 and 4)
while they suffer head wind in the other sub-Keplerian regions.
In the result of Figure~\ref{fiducial}, the super-Keplerian region
exists at $0.0<x/H \la 2.0$ in the quasi-steady state.

\subsection{Dependence on widths of stable and unstable regions}

In the limit of $L_{\rm s} \rightarrow 0$ or $L_{\rm u} \rightarrow \infty$, 
the system would have uniform turbulence and
would not acquire the local rigid rotation.
To derive the condition for establishment of the local rigid rotation, 
we perform runs with different values of $L_{\rm s}$ and $L_{\rm u}$.

\begin{figure}[tbp]
\begin{center}
\includegraphics[width=100mm]{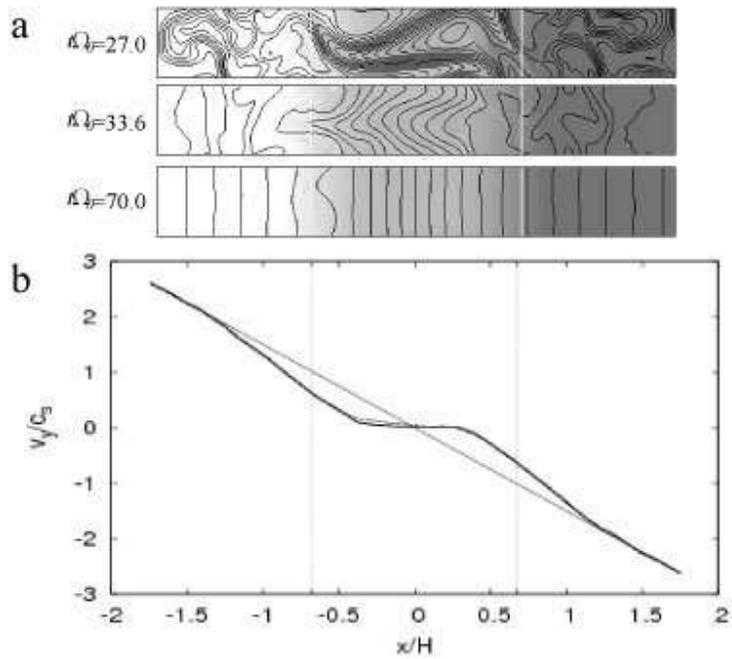}
\end{center}
\caption{Results of model-s11. 
(a) Time evolution of the magnetic field (solid lines) and 
angular velocity $v_y$ (contours) on the $x$-$z$ plane.
(b) Time evolution of vertically averaged 
angular velocity ($v_y$).  
The bold, dashed and thin solid lines express
the snapshots at $t\Omega_{0} = 0.0, 40.0$ and $70.0$,
respectively.
The meanings of lines are
the same as Fig.~\ref{fiducial}.
}
\label{s1}
\end{figure}

In the fiducial case, $L_{\rm u}=1.4H$ and $L_{\rm s}=4.0H$.
We first present the results of a series of runs with 
various values of $L_{\rm s}$ and the fixed $L_{\rm u}$. 
In model-s11, model-s055, and model-s005, 
$L_{\rm s} = 1.1H$.  
$0.55H,$ and $0.05H$, respectively. 
The other parameters are the same as those in the fiducial model.
Figure~\ref{s1} describes the evolution of the magnetic field in model-s11. 
Since the stable region is narrower than the fiducial case, 
the magnetic perturbations that arise in the unstable region
propagate to the boundary of the computational box
before they are dissipated.
The whole region including the initially stable region becomes temporarily turbulent.
However, the perturbations that reach the boundary are
weakened and do not have enough momentum to go into the unstable region again.
The turbulence ceases after $t\Omega_{0} \ga 40$ and
the quasi-steady rigid rotation is obtained as the result, although
the pressure contrast is smaller than that in the fiducial model
(the minimum pressure is $\simeq 0.65P_0$, while it is
$\simeq 0.33 P_0$ in the fiducial model).

\begin{figure}[tbp]
\begin{center}
\includegraphics[width=100mm]{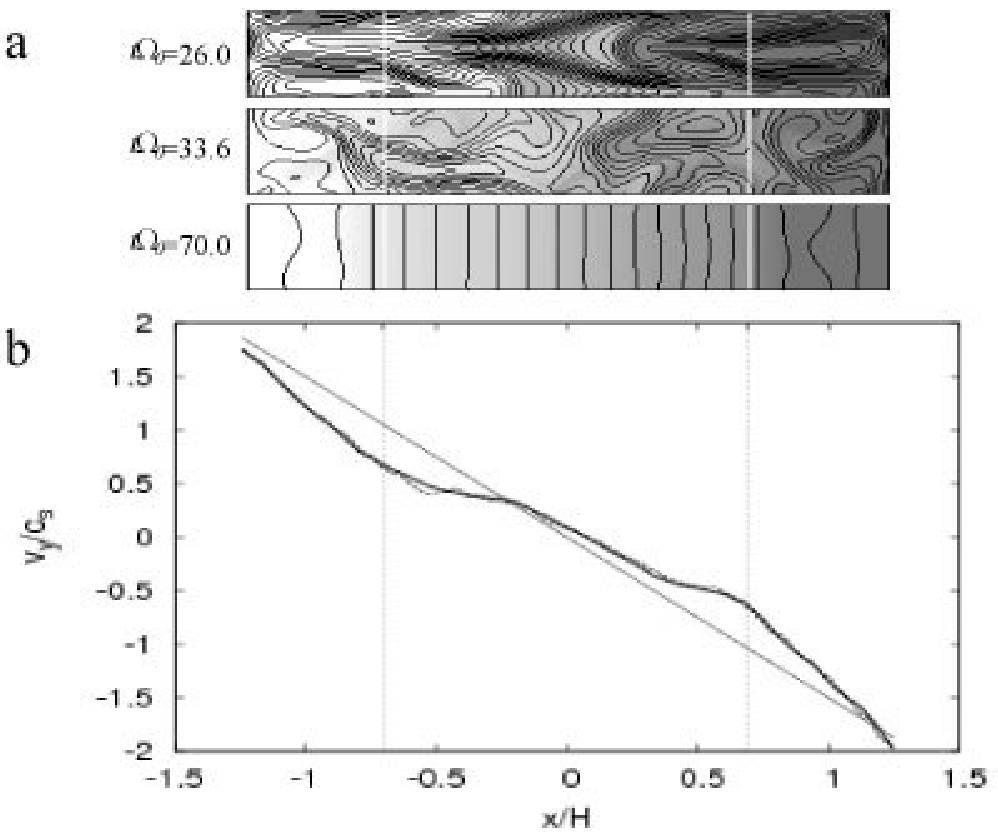}
\end{center}
\caption{
The same plots as Fig.~\ref{s1} except for model-s055.
}
\label{s05}
\end{figure}

The evolution of the MRI in model-s055 is shown in Figure~\ref{s05}. 
The narrower stable region allows the magnetic perturbations to 
come back to the unstable region because of the periodic boundary condition. 
The turbulence lives longer, but it eventually vanishes 
by $t\Omega_{0} \sim 70$.
The quasi-steady velocity profile which is more flattened than the Keplerian is established, although it is not as distinctive as the rigid rotation
in the fiducial model.

\begin{figure}[tbp]
\begin{center}
\includegraphics[width=60mm]{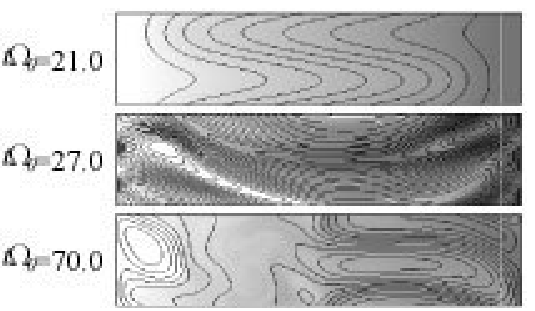}
\end{center}
\caption{
The same plots as Fig.~\ref{s1}a except for model-s005.
}
\label{s005}
\end{figure}

Figure \ref{s005} shows the result of model-s005.
Because of the small stable regions, turbulence expanded to the entire
region does not cease and uniform turbulence is maintained.
Efficient angular momentum transfer in the entire region prevents
velocity field from having a quasi-steady super-Keplerian region.

\begin{figure}[tbp]
\begin{center}
\includegraphics[width=90mm]{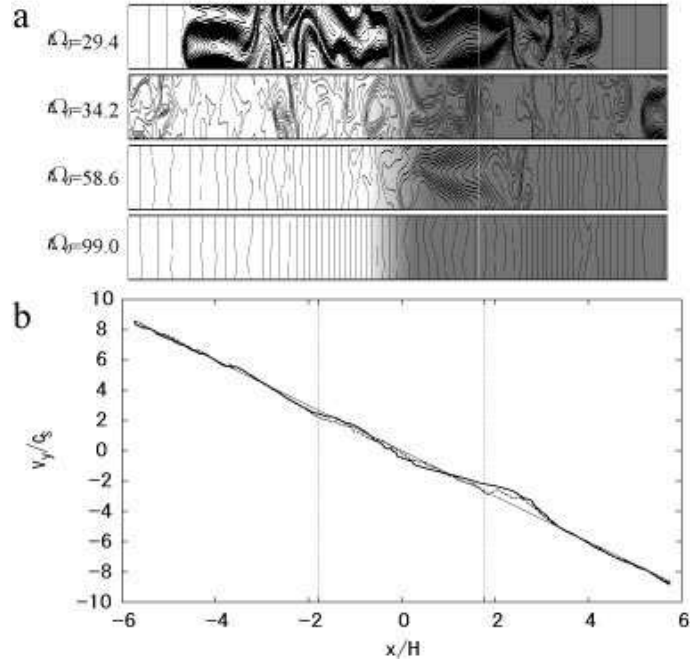}
\end{center}
\caption{
The same plots as Fig.~\ref{s1} except for model-u34.
}
\label{u34}
\end{figure}

So far, we have changed values of $L_{\rm s}$ while $L_{\rm u}$ is constant.
We also performed a run (model-u34) with $L_{\rm u}=3.4H$ that is 
enlarged from the fiducial model.
The other parameters are the same as those in the fiducial model. 
The results are shown in Figure~\ref{u34}.
In the enlarged unstable region, the magnetic field is stretched enough for reconnection.
After the reconnection, long-lived magnetic perturbations are sustained
around the boundary between unstable and stable regions 
(the panel at $t\Omega_{0}=58.5$). 
The magnetic perturbations reach the computational boundary, because 
the conversion from magnetic energy to kinetic energy 
during the magnetic reconnection adds horizontal motion to the fluid element. 
However, the magnetic perturbations are not strong enough 
to pass through the stable region
and come back to the unstable region. 
As a result, the evolution is similar to model-s11, except that the
rigid rotation is established in the regions 
near the boundary between unstable and stable regions.
The panels at $t\Omega_{0}=61.5$ and $t\Omega_{0}=99.0$ show that
this flow pattern is also quasi-steady.

\begin{figure}[tbp]
\begin{center}
\includegraphics[width=90mm]{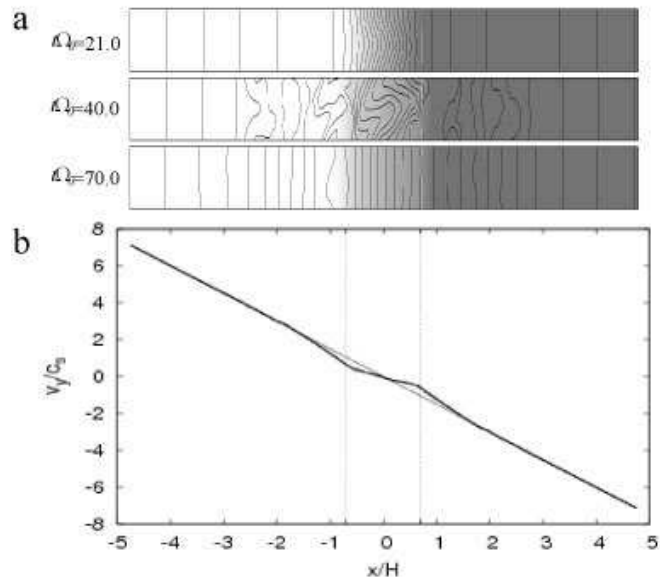}
\end{center}
\caption{
The same plots as Fig.~\ref{s1} except for model-$\eta 0$.
}
\label{eta}
\end{figure}

These results suggest that dissipation of magnetic perturbations
during the passage through stable regions plays an
essential role in creation of rigid rotation regions.
We performed additional series of runs with 
a larger value of the resistivity, $\eta=0.0028\Omega_{0}H^2$.
The result of model-$\eta0$ with $L_{\rm u}=1.4H$ and $L_{\rm s}=4.0H$ (the same as
the fiducial model) is shown in Figure~\ref {eta}.
Because of the faster dissipation, the quasi-steady rigid rotation is
established more early, although the weaker angular momentum transfer leads to
smaller pressure contrast in the quasi-steady state.
Even if we reduce $L_{\rm s}$ to $\sim 0.5H$, 
the magnetic perturbations do not arrive at the boundary.
We also performed runs with further larger values of 
the resistivity, $\eta/(\Omega_{0}H^2) = 0.0012, 0.0036, 0.0044, 0.0052$
and 0.0060 (Table 1) to confirm that the quasi-steady rigid rotation 
is always established for these values of $\eta$.
We discuss the dependence on the resistivity in the next subsection.

\subsection{Condition for the rigid rotation}\label{type}

As we have shown, establishment of the quasi-steady rigid rotation depends
on $L_{\rm u}$, $L_{\rm s}$, and $\eta$.
Through the results with various $L_{\rm u}$, $L_{\rm s}$, and $\eta$
as listed in Table 1, we found that the results are 
summarized by a single parameter,
a spatially averaged magnetic Reynolds number, defined by \citep{sam99, sai04}
\begin{equation}
R_{\rm m,ave}=\frac{v^2_{{\rm A}z,{\rm ave}}}{\eta\Omega_{0}},
\end{equation} 
where $v_{{\rm A}z,{\rm ave}}$ is evaluated by
a spatially averaged vertical magnetic field ($B_{z,{\rm ave}}$)
at the initial stage.
In our simulation setting, $B_{z,{\rm ave}} \simeq 
B_0(L_{\rm u} + 2 \cos 85^{\circ} L_{\rm s})/(L_{\rm u} + 2 L_{\rm s})$.

In many cases, the initial distributions of vertical magnetic field 
are spatially smoothed out by turbulent diffusion after $\sim 10$ orbits.
If the remaining magnetic field ($\sim B_{z,{\rm ave}}$) 
is still large enough to globally cause MRI turbulence, 
the transition to the quasi-steady state
does not occur, as seen in the results of model-s005.
Linear theory shows that the critical magnetic Reynolds number
for occurrence of MRI is $R_{\rm m} \sim 1$ \citep{sam99, sai04}.
Thus, $R_{\rm m,ave}$ regulates the establishment of the quasi-steady 
rigid rotation in our simulations.

\begin{figure}[tbp]
\begin{center}
\includegraphics[width=100mm]{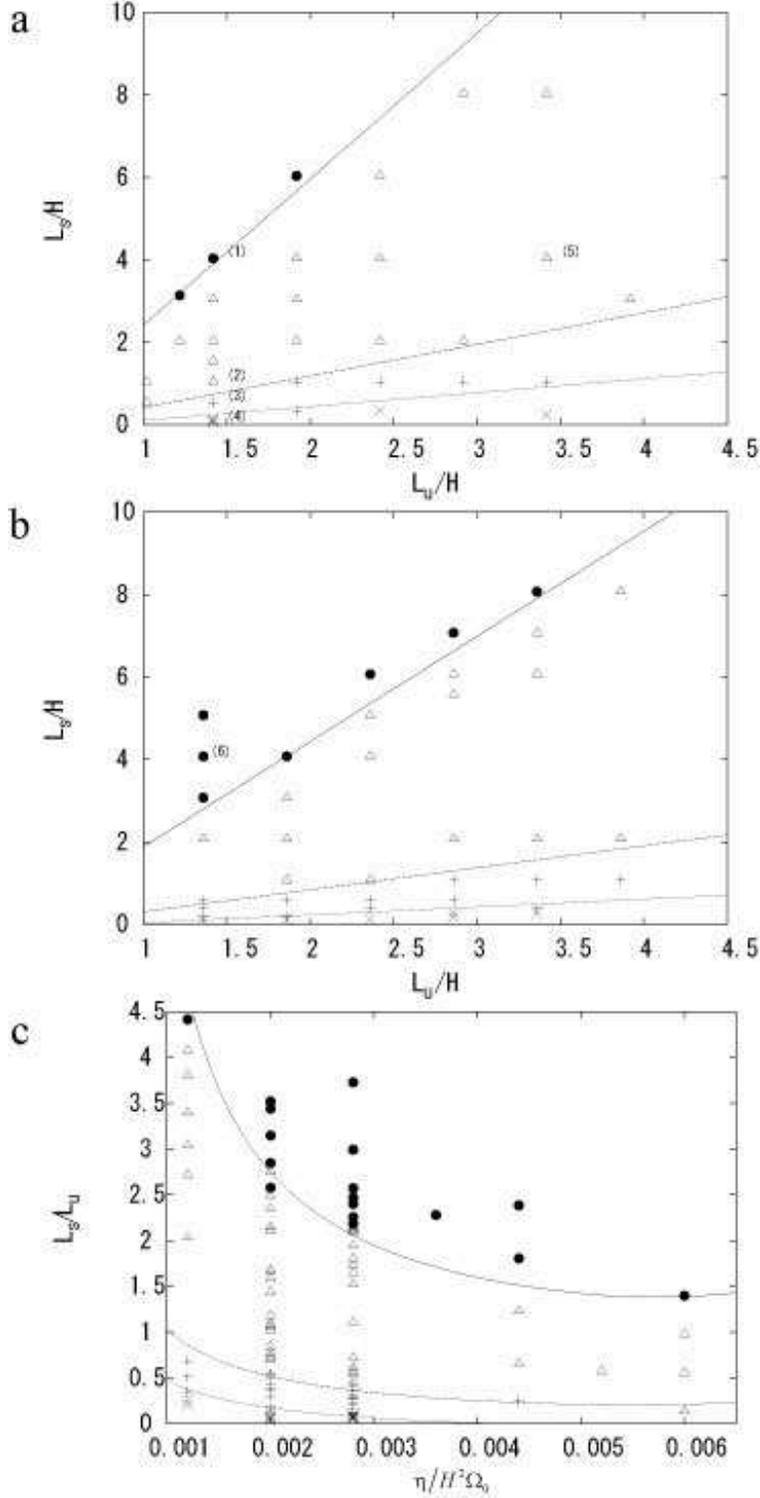}
\end{center}
\caption{
The classification of results by the magnetic Reynolds number. 
Runs with (a)$\eta=0.0020\Omega_{0}H^2$ and (b)$\eta=0.0028\Omega_{0}H^2$ 
on the $L_{\rm u}-L_{\rm s}$ plane 
and (c)all the runs on the $\eta-L_{\rm s}/L_{\rm u}$ plane. 
The four types of results, A, B, C and D, are represented by 
filled circles, triangles, daggers and crosses. 
The solid, dashed and dotted lines express $R_{\rm m,ave}=0.1, 
0.5$ and $1.0$ respectively. 
The runs with numbers express 
(1)fiducial model; (2)model-s11; (3)model-s054; (4)model-s005
(5)model-u34; (6)model-$\eta 0$.
}
\label{class}
\end{figure}

We summarize the results of 91 runs with different parameters 
in  Figure~\ref{class} and found that evolution of the magnetic 
field is indeed classified into four types by the values of 
$R_{\rm m,ave}$ as follows:
\begin{description}
\item[Type A ($R_{\rm m,ave} \la 0.1$):]
Local MRI turbulence generated in the initially unstable region
propagates both inward and outward, but
the magnetic perturbations are dissipated before they reach 
the boundary of the simulation box.
After a few tens of orbits, the turbulence
vanishes in the entire region and the quasi-steady flow
is established.
In the initially unstable region, rigid rotation flow is 
resulted in by angular momentum transfer due to
the MRI turbulence. 
This class is represented by filled circles in Figure~\ref{class} 
and it includes the fiducial model and model-$\eta0$.

\item[Type B ($0.1\la R_{\rm m,ave} \la 0.5$):] 
The magnetic perturbations reach the boundary, but they do not 
intrude back into the original unstable region. 
The quasi-steady rigid-rotation region appears as in Type A.
This class is represented by triangles in Figure~\ref{class} 
and it includes model-s11 and model-u34.

\item[Type C ($0.5 \la  R_{\rm m,ave} \la 1.0$):] 
The magnetic perturbations intrude the unstable region 
after the passage through the stable regions. 
However, the diffused $B_{z}$ is not large enough to globally maintain
turbulence and the quasi-steady rigid rotation region is still formed, 
although their locations are not necessarily the same as in Type A and B.
This class is represented by daggers in Figure~\ref{class} 
and it includes model-s055.

\item[Type D ($1.0 \la R_{\rm m,ave}$):] 
Even after the turbulent diffusion, $B_{z}$ is able to maintain
the turbulence in the entire region.
Because of the uniform turbulent state, 
the quasi-steady rigid rotation state is not generated. 
This class is represented by crosses in Figure~\ref{class} 
and it includes model-s005.

\end{description}

\section{Conclusion and Discussion}\label{discuss}

We have investigated evolution of 
patchy magneto-rotational instability (MRI) due to radially
non-uniform magnetic field and found that, under some conditions,
the original Keplerian shear flow is transformed into
quasi-steady profile involving a local 
rigid-rotation regions.
The outer parts of the rigid-rotation regions
are generally super-Keplerian.
Such a situation would arise in the outer boundary of
MRI dead zone as well as the inner boundary in a protoplanetary disk, 
as discussed in section 1. 

Assuming uniformity in the azimuthal direction of disks,
we have carried out two-dimensional resistive MHD simulations 
in a shearing box model with periodic boundary conditions.
We set up both stable and unstable regions in the box,
changing direction of the vertical seed magnetic field ($B_z$) non-uniformly.
In the initially unstable region,
MRI turbulence is generated locally and magnetic perturbations
propagate both radially inward and outward by the turbulent diffusion. 
If the unstable region is sufficiently large compared with 
the stable region, the turbulence eventually covers the entire
region and the initial non-uniformity vanishes.
However, if the stable region is relatively large, diffused
magnetic perturbations no more maintain MRI turbulence.
After the turbulence ceases,
the initial flow of uniform Keplerian shear is transformed 
into a different quasi-steady state.
In the quasi-steady state, rigid-rotation is established locally.
The deviation from Keplerian shear motion is supported 
by pressure gradient that has been produced also by mass transport
associated with the tentative turbulence. 
Through simulations with various initial conditions, we found 
that the quasi-steady rigid rotation is established
if the spatially averaged 
magnetic Reynolds number satisfies $R_{\rm m,ave} \la 1$
in the initial state.

Because the center of the local rigid rotation is often
Keplerian, super-Keplerian flow 
appears in the outer parts of the rigid rotation region.
As explained in section 1, dust grains and planetary embryos 
can be trapped in the boundary between regions of 
sub- and super-Keplerian motion
through radial migration induced by aerodynamic gas drag
and type I migration. 
The boundary is coincident with pressure maxima in the quasi-steady state. 

The effect of global pressure gradient 
is included by shifting the initial gas velocity 
from the pure Keplerian speed to slight sub-Kepler.
The sift, which is the velocity difference between the disk gas 
and the dust grains, is
\citep[e.g.,][]{adachi76,wei77}
\begin{equation}
\Delta v_y = \frac{c_s^2}{2 v_{\rm K}^2} \frac{d \ln P}{d \ln r} v_{\rm K}
\simeq - 5 \times 10^{-2} \left(\frac{r}{\rm 1AU} \right)^{3/4} c_s
\end {equation}
where $d \ln P/d \ln r$ is the global pressure gradient and
the temperature distribution in the limit of optically thin disks 
around solar-luminosity stars, $T = 280 (r/1{\rm AU})^{-1/2}$, is assumed.
Since maximum values of $\Delta v_{y}$ in the super-Keplerian regions
are $\ga 0.4 c_s$ in our results, 
the MRI effects can easily surpass the global pressure gradient effects 
and super-Keplerian regions can emerge even when the initial down-shift 
in the gas rotation velocity is present.

If the super-Keplerian region is sustained long enough for 
dust grains to accumulate, planetesimals can be formed through
self-gravitational instability \citep{you02,joha06}. 
In the case of $R_{\rm m,ave} \la 1$ in which MRI turbulence
ceases in the entire region after a few tens of orbits, 
we found that $B_z$ is still large enough in the rigid-rotation region.
In that region, MRI is suppressed by disappearance of shear motion
but not by dissipation of $B_z$ (in the regions other than
the rigid-rotation region, diffused-out $B_z$ is smaller than
the value for MRI to occur).
If the rigid rotation tries to go back to the original
Keplerian shear motion, MRI turbulence again occurs 
and it transfers angular momentum to recover the rigid rotation.
Thus, the rigid rotation and hence the associated super-Keplerian 
rotation are self-sustaining.
When the remaining $B_z$ is diffused out by ohmic dissipation
on timescale $\sim L_u^2/\eta \sim 10^3 \Omega_0^{-1}$, such
stabilization mechanism is no more effective.
Then, the rigid rotation can go back to the original
Keplerian shear motion by the residual uniform viscosity.
However, since MRI no more occurs, the residual viscosity 
would be very small.
Thus, it is expected that the super-Keplerian regions
would survive long enough for accumulation of dust grains and
formation of planetesimals.
We also did a calculation starting from the end result of the
super-Keplerian rotation state, artificially modifying $B_z$ to
uniform distribution.
However, we do not see any relaxation of the velocity field
back to Keplerian rotation within the timescales of 40 orbits.

In the next paper, we will demonstrate the accumulation of dust grains and discuss the effect of the azimuthal magnetic field especially in stable region by three-dimensional simulation. 
We will also show the results of non-uniform resistivity case 
with constant $B_{z}$, which may be more likely 
to occur at the outer edge of a dead zone. 
We find similar appearance of super-Keplerian regions,
because intrinsic physics to
transform initial Keplerian flow to
quasi-steady non-uniform rotation flow 
is temporal generation of MRI and the stabilization of the MRI 
due to the established
rigid rotation but not due to dynamo of magnetic field (see section 3).

The appearance of the super-Keplerian region
also halts inward type I migration of planetary
embryos \citep{tan02,mass06}.
Since the process we found also works at the outer
boundary of a dead zone and the outer boundary
migrate from $\sim 10$AU to the proximity of the central star,
this process may also help the formation of cores massive enough
to onset runaway gas accretion and retain terrestrial planets
against type I migration.
This may play an important role in frequency of 
extrasolar gas giants and habitable planets.
We will address this issue with sequential planet formation
model \citep{idalin04,idalin08a,idalin08b} in a future paper.

We thank detailed helpful comments by an anonymous referee.

\appendix
\section*{Appendix 1. Resolution test}

\begin{figure}[tbp]
\begin{center}
\includegraphics[width=200mm]{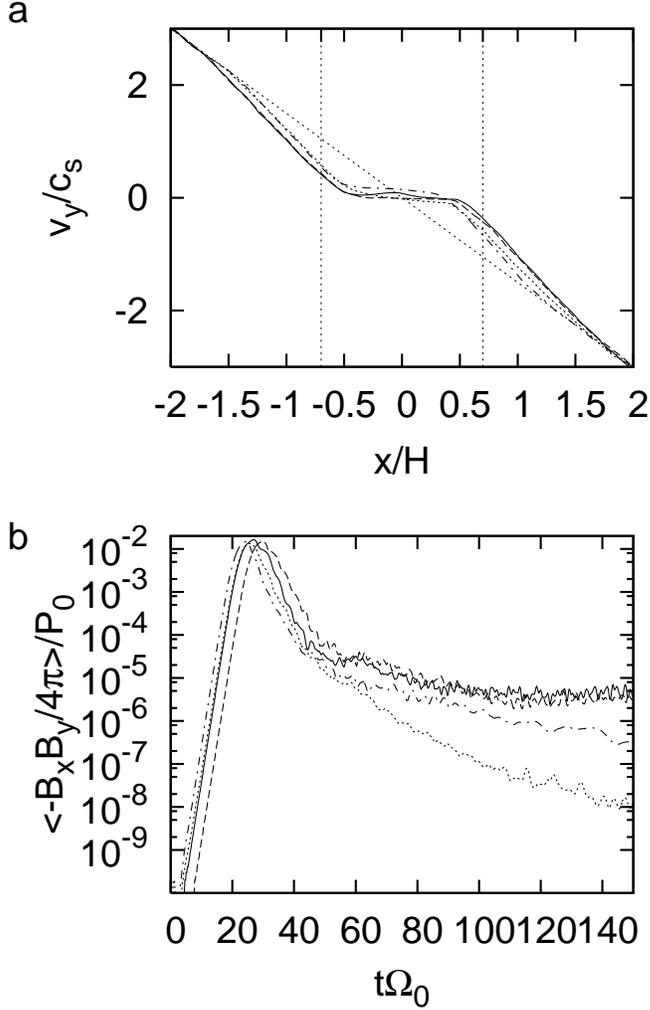}
\end{center}
\caption{
Results for various resolutions. 
(a)Close-up snapshots of vertically averaged angular velocity($v_y$) 
at $t\Omega_{0}=150.0$. 
(b)Time evolution of the volume-averaged Maxwell stress which is normalized by 
the initial pressure $P_{0}$. 
Resolutions corresponding to $dx=dz=0.005H$, $0.01H$, $0.0156H$ and $0.025H$ are represented by dashed, bold, dash-dotted and dotted lines respectively.
}
\label{resolution}
\end{figure}

We have investigated the effects of numerical resolution on our results 
using our fiducial model. Four cases are studied. The distribution of 
angular velocity and the average Maxwell stress are shown in Figure 
~\ref{resolution}. The most important issue in our results is the emergence 
of the quasi-steady state in which sub- and super-Keplerian areas exist. 
Figure~\ref{resolution}a shows this to be seen even in the worst resolution 
case. In addition, the Maxwell stress in Figure~\ref{resolution}b appears to 
be converging at resolutions above $dx=0.01H$ while this quantity tends to 
increase with resolution.

\citet{fro07} showed the decrease of Maxwell stress with increasing resolution 
and stated that the MRI turbulence activity is notoriously ill-behaved in high 
resolution calculation. Our resolution test, however, shows the convergence 
with increasing resolution. This could be because the ohmic dissipation 
(resistivity) is kept constant in our resistive MHD simulations while the 
simulation of \citet{fro07} included only the numerical resistivity and it 
deacreases with increasing resolution. However, more detailed study is needed 
to clarify the difference in convergence between our simulation and the 
simulation of \citet{fro07}, which is left to our future study. 

This fact will be 
important when we investigate the motion of dust particles. The degree of 
particle concentration may be depend on the turbulence activity. While the 
details are left for a future study, one may reasonably expect the 
possibility of dust concentration at the outer-edge of super-Keplerian area.

Figure~\ref{resolution}b also shows the eligibility of the integration time.
The saturation level varies only slightly in $t\Omega_{0}>60.0$
in the two high resolution cases. The quasi-steady state has already been
created by this time. These indicate that our choice the integration time 
$t\Omega<100.0$
is validated.

\section*{Appendix 2. Dispersion relations}

Our simulations show that the MRI stabilization in strong magnetic field
with nearly rigid rotation, which is consistent with the linear analyses.
The linear analysis using ideal MHD equations 
in \citet{bal91} gave the critical wavelength (equation (2.14b) 
in the paper, with neglect of the Brunt-V\"{a}is\"{a}l\"{a} frequency):
\begin{equation}
\left|\lambda_{z,{\rm crit}}\right| = v_{{\rm A}z}
\left|\frac{d\Omega^{2}}{d\ln r}\right|^{-1/2} 
=\frac{v_{{\rm A}z}}{\Omega} (2\left|q\right|)^{-1/2}, 
\label{linear-ideal} 
\end{equation}
where $q$ is defined as $\Omega(r) \propto r^{-q}$. 
The perturbations with wavelength shorter than $\lambda_{z,{\rm crit}}$ are stable.
When the rotation becomes rigid rotation ($q \rightarrow 0$), 
$\lambda_{z,{\rm crit}}$ becomes large.
If $\lambda_{z,{\rm crit}}$ is larger than the scale height of a disk, 
the system is stable, irrespective of magnetic field strength. 

\begin{figure}[tbp]
\begin{center}
\includegraphics[width=200mm]{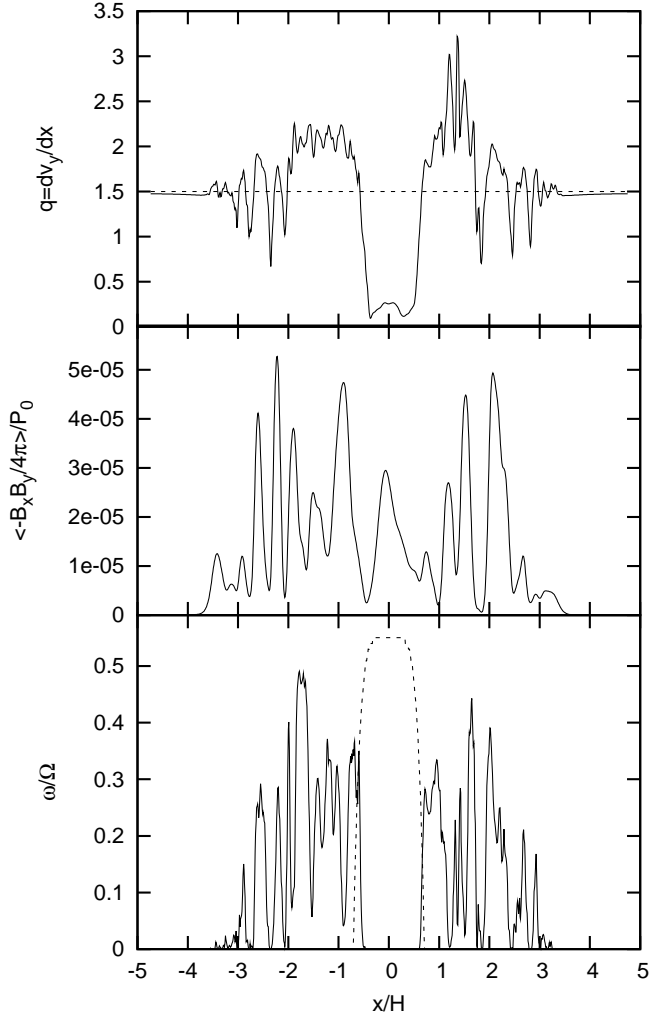}
\end{center}
\caption{
Temporally and vertically averaged shear rate $q=-dv_{x}/dt$ ({\it top}), 
Maxwell stress normalized by initial pressure 
$- \langle B_{x}B_{y} \rangle /4\pi P_{0}$ ({\it middle}) and the 
growth rate estimated by the dispersion relation 
(equation (2) in Appendix 2) with the simulated values ({\it bottom}) 
of the fiducial model. Bold lines are values temporally averaged 
over $t\Omega_{0}=45.0-70.0$ and dashed lines mean the initial values.
}
\label{growth}
\end{figure}

With the effect of ohmic dissipation,
the dispersion relation was obtained by 
\citet{jin96} and \citet{sam99} as
\begin{equation}
\sigma^{4}+2\xi\sigma^{3}
+\left(2q_{z}^2+\xi^{2}+\frac{\kappa^{2}}{\Omega^2}\right)\sigma^{2}
+2\xi\left(q_{z}^{2}+\frac{\kappa^{2}}{\Omega^{2}}\right)\sigma
+\left(-4+\frac{\kappa^{2}}{\Omega^{2}}\right)q_{z}^{2}
+q_{z}^{4}+\frac{\kappa^{2}}{\Omega^{2}}\xi^{2}=0,
\label{linear-eta}
\end{equation}
where $\sigma$ is a growth rate in units of orbital frequency, 
$q_{z}=k_{z}v{{\rm A}z}/\Omega$, $\xi=k_{z}^{2}\eta/\Omega$, and
$\kappa$ is an epicyclic frequency defined by
\begin{equation}
\kappa^{2}=\frac{2\Omega}{r}\frac{d\left(r^{2}\Omega\right)}{dr}
=\left(2-q\right)2\Omega^{2}. 
\end{equation}
This dispersion relation is derived with 
the assumption of uniform density and negligible 
Brunt-V\"{a}is\"{a}l\"{a} frequency. 
Since the density is almost uniform in the rigid rotation in our simulation,
we apply this dispersion relation to calculate the predicted $\sigma$
with the quantities obtained by our simulation.
We plot the temporally and vertically averaged $q$ ($=-dv_{x}/dx$), 
Maxwell stress ($-\langle B_{x}B_{y} \rangle/4\pi P_{0}$) 
and the evaluated growth rate $\sigma$ in Figure~\ref{growth}. 
The growth rate is very small in the middle region of nearly rigid rotation 
($q \ll 1$), although Maxwell stress is not small there.
Thus, we conclude that MRI is suppressed by
established nearly rigid rotation, but not by dissipation of magnetic field.



\clearpage

\clearpage
\clearpage
%
%
%
%
%
%
%
%
%
%
%
%
\clearpage
\clearpage
\begin{table}
\begin{tabular}{llllll}
\hline \hline
$\eta/\Omega_{0}H^2$ & $L_{\rm u}/H$ & $L_{\rm s}/H$ & 
$R_{\rm m,ave}$ & {\rm result} & $\Delta v_{y}/c_s$ \\ \hline

0.0012 & 1.5 & 6.5, 6.0, 5.0, 4.0 & 0.10, 0.11, 0.13, 0.16 
             & A, B, B, B & 0.82, 1.2, 0.91, 0.86 \\
       &     & 1.0, 0.51 & 0.61, 1.1 & C, C & 0.66, 0.51\\
       & 2.0 & 7.5, 6.0, 4.0 & 0.13, 0.16, 0.23 & B, B, B 
             & 1.7, 1.1, 0.98 \\
       &     & 1.0, 0.51 & 0.89, 1.4 & C, D & 0.68, ... \\
       & 2.5 & 0.51 & 1.7 & D  & ... \\ \hline
0.002 & 1.0 &  1.1, 0.54 & 0.23, 0.42 & B, B  & 0.40, 0.38 \\
      & 1.2 &  3.1       & 0.10       & A     & 0.57 \\
      & 1.4 &  $4.0^{1}$, 3.1, $1.1^{2}$, $0.55^{3}$ 
            & 0.096, 0.13, 0.37, 0.64 
            &  A, B, B, C & 0.73, 0.55, 0.48, 0.41 \\
      &    &  0.15, 0.10, $0.05^{4}$ & 1.2, 1.4, 1.5 
           &  D, D, D  & ..., ..., ... \\
      & 1.9 &  6.0, 4.0, 3.0 & 0.094, 0.14, 0.19
           &  A, B, B  & 0.97,0.88, 0.54 \\
      &    &  2.0, 1.0, 0.35 & 0.28, 0.53, 1.1
           & B, C, C & 0.87, 0.72, 0.43 \\
      & 2.4 &  2.0, 0.35 & 0.38, 1.3 & B, D  & 1.2, ... \\
      & 2.9 & 10.0, 8.0, 1.0 & 0.092, 0.11, 0.81 & A, B, C  
           & 1.2, 1.4, 0.99 \\
      & 3.4 & 12.0, 8.0, $4.0^{5}$, 0.24 & 0.090, 0.14, 0.28, 1.8  
           & A, B, B, D& 1.0, 1.2, 1.1, ... \\
      & 3.9 &  3.1 &  0.44 & B  & 0.77 \\ \hline 
0.0028 & 1.4 &  5.1, $4.0^{6}$, 2.1, 0.57 & 0.15, 0.069, 0.14, 0.45
             & A, A, B, C  & 0.35, 0.45, 0.28, 0.23 \\
       &     &  0.37, 0.17, 0.11 & 0.61, 0.87, 1.00 & C, C, D
            & 0.34, 0.14, ... \\
       & 1.9 & 4.0, 3.1, 2.1, 1.1 & 0.099, 0.14, 0.20, 0.38
            & A, B, B, B  & 0.48, 0.31, 0.49, 0.42 \\
       &     &  0.57, 0.17, 0.11 & 0.61, 1.05, 1.2 & C, C, D
            & 0.40, 0.25, ... \\
       & 2.4 & 6.1, 5.1, 4.0, 1.1 & 0.088, 0.11, 0.13, 0.49
            & A, B, B, B  & 0.50, 0.62, 0.66, 0.53 \\
       &     & 0.57, 0.37, 0.11 & 0.75, 0.92, 1.25 & C, C, D
            & 0.46, 0.28, ... \\
       & 2.9 & 7.1, 6.1, 5.1, 2.1 & 0.093, 0.11, 0.12, 0.47
            & A, B, B, B  & 0.60, 0.42, 0.50, 0.48 \\
       &     & 1.1, 0.57, 0.27, 0.17 & 0.58, 0.85, 1.13, 1.26 & C, C, D, D
             & 0.56, 0.37, ..., ... \\
       & 3.4 & 8.1, 7.1, 6.1, 2.1 & 0.0097, 0.11, 0.13, 0.39
             & A, B, B, B  & 1.00, 0.79, 0.58, 0.78 \\
       &      & 1.1, 0.37, 0.27 & 0.66, 1.1, 1.3 & C, C, D
             & 0.48, 0.61, ... \\
       & 3.9 & 8.1, 2.1, 1.1 & 0.11, 0.45, 0.73
             & B, B, B  & 0.46, 0.43, 0.49 \\ \hline 
0.0036 & 1.8 &  4.1 &  0.077 & A  & 0.28 \\
0.0044 & 1.7 &  4.1, 3.1 &  0.063, 0.085
             & A, A  & 0.17, 0.18\\
       &     &  2.1, 1.1, 0.43 & 0.13, 0.24, 0.50
             & B, B, B  & 0.20, 0.18, 0.18 \\
0.0052 & 2.3 &  1.3 & 0.24 & B  & 0.15 \\
0.0060 & 2.4 &  3.3, 2.3, 1.3, 0.32 
       & 0.084, 0.36, 0.21, 0.50 & A, B, B, B  & 0.22, 0.17, 0.21, 0.20 \\
\hline
\end{tabular}
\end{table}
%
%
%
\begin{table}
\caption{Simulation parameters and results for 91 runs.
$\eta$, $L_{\rm u}$, $L_{\rm s}$, $R_{\rm m,ave}$,
and $\Delta v$ are 
magnetic resistivity, radial width of unstable region, 
that of stable regions, initial averaged magnetic Reynolds number
and the maximum deviation from Kepler velocity, respectively.
For highly turbulent cases (marked ``D''), $\Delta v_{y}$ is omitted.
The fifth column ``result'' indicates classification of 
the results (\S 3.3).
Multiple values in the columns correspond to different runs.
(1) fiducial model;
(2) model-s11; (3) model-s055; (4) model-s005; 
(5) model-u34; (6) model-$\eta 0$.
}
\label{tab:1}
\end{table}
\end{document}